# Extending species-area relationships (SAR) to diversity-area relationships (DAR)[©]

Zhanshan (Sam) Ma
Computational Biology and Medical Ecology Lab
State Key Lab of Genetic Resources and Evolution
Kunming Institute of Zoology
Chinese Academy of Sciences
Kunming 650223, China
samma@uidaho.edu

## Abstract

**Aim**: I extend the traditional SAR, which has achieved status of ecological law and plays a critical role in global biodiversity assessment, to the general (alpha- or beta-diversity in Hill numbers) diversity area relationship (DAR). The extension was motivated to remedy the limitation of traditional SAR that only address one aspect of biodiversity scaling—species richness scaling over space.

**Innovation**: The extension was made possible by the fact that all Hill numbers are in units of species (referred to as the *effective number of species* or as *species equivalents*), and I postulated that Hill numbers should follow the same or similar pattern of SAR. I selected three DAR models, the traditional power law (PL), PLEC (PL with exponential cutoff) and PLIEC (PL with inverse exponential cutoff). I defined three new concepts and derived their quantifications: (*i*) DAR profile—$z$-$q$ series where $z$ is the PL scaling parameter at different diversity order ($q$); (*ii*) PDO (pair-wise diversity overlap) profile—$g$-$q$ series where $g$ is the PDO corresponding to $q$; (*iii*) MAD (maximal accrual diversity) profile—$D_{max}$-$q$ series where $D_{max}$ is the MAD corresponding to $q$. Furthermore, the PDO-$g$ is quantified based on the self-similarity property of the PL model, and $D_{max}$ can be estimated from the PLEC parameters. The three *profiles* constitute a novel DAR approach to biodiversity scaling.

**Main conclusions**: I verified our postulation with the American gut microbiome project (AGP) dataset of 1473 healthy North American individuals (the largest human dataset from a single project to date). The PL model was preferred due to its simplicity and established ecological properties such as self-similarity (necessary for establishing PDO profile), and PLEC has an advantage in establishing the MAD profile. All three profiles for the AGP dataset were successfully quantified and compared with existing SAR parameters in the literature whenever possible.

**Running Title**: Extending SAR to DAR

---

[©] Ma ZS (2017) Extending SAR to DAR. preprint posted on https://arxiv.org/

# Introduction

The species area relationship (SAR), well regarded as one of the few laws in ecology and biogeography that has been pursued by generations of ecologists and biogeographers since the 19th century (Watson 1835). SAR relationship is thought to have inspired MacArthur and Wilson's (1967) island biogeography theory and is even suggested to play an important role in setting conservation strategy and polices. Extensive theoretical, empirical and application studies on SAR have been performed in macro-ecology in the last several decades, which establishes SAR as one of the few most important ecological laws (*e.g.*, Preston 1960, Connor & McCoy 1979, Rosenzweig 1995, Lomolino 2000, Tjørve 2008, 2009, Drakare et al. 2006, Harte et al. 2009, He & Hubbell 2011, Sizling et al. 2011, Storch et al. 2012, Triantis et al. 2012, Helmus et al. 2014). It is hailed as "ecology's most general, yet protean pattern" by Lomolino (2000) and Whittaker & Triantis (2012). Although the study of SAR originated in macro-ecology of the plants and animals, thanks to the revolutionary genomic and especially metagenomic sequencing technologies, microbial ecologists have already joined in the exploration starting approximately a decade ago (Green et al. 2004, Horner-Devin et al. 2004, Bell et al. 2005, Noguez 2005). The metagenomic revolution originated in environmental microbiology, and later expanded to the human microbiome research through the launch of the US NIH human microbiome project (HMP, apparently a natural follow-up of the HGP since human metagenome is often referred as the *second* human genome) around 2007 (HMP Consortium 2012, Turnbaugh et al. 2007), and more recently expanded to the earth microbiome project (EMP) (Gilbert 2011). Indeed, the ecological theory has been both a unifying driving force and testbed for this revolution (*e.g.,* Fierer (2008), Costello et al., 2012, Lozupone et al. 2012, Haegeman et al. 2013, Barberán et al. 2014, Ma et al. 2012, Ma 2015, Chiu & Chao 2015). Today, ecologists are capable more than ever to test major ecological theories across not only taxa (plants, animals and microbes) but also ecosystem types (*e.g.,* forest, lakes, ocean, human and animal microbiomes), and novel findings and insights are revealed more frequently than ever. For example, a recent study published by Sunagawa et al. (2015) in the journal *Science* found that the global ocean microbiome and human gut microbiome share a functional core of prokaryotic gene abundance with 73% in the ocean and 63% in the gut, in spite of the huge physiochemical differences between the two ecosystems.

Methodologically, I extend the traditional SAR models to the diversity-area relationship (DAR) models by using the Hill Numbers as a general diversity metric system, including both *alpha* and *beta* diversities. The choice of Hill numbers not only makes our study more comprehensive than existing SAR studies, but also makes it possible to extend the traditional SAR to both alpha and beta-diversities beyond species richness (*i.e.,* beyond the zero-order diversity in terms of the Hill numbers). I take advantage of the recent consensuses that Hill numbers offer the best measure of alpha diversity and that multiplicative beta-diversity partition of the Hill numbers is more appropriate than the additive partition (Jost 2007, Ellison 2010, Chao et al. 2012, 2014). For example, rather than building a SAR model, I present DAR models for a series of diversity orders from the zero-order (richness), first order (a function of Shannon index), second order (a function of Simpson index), third and fourth orders, resulting a *diversity-area scaling profile* (DAR profile), corresponding to the diversity profile of Hill numbers (Chao et al. 2012, 2014).

In consideration of the debates on the functional forms of SAR (*e.g.*, He & Hubbell 2012), I tested the traditional power law model (PL) as well as what I believe are two most promising extensions, *i.e.*, the PL with exponential cutoff (PLEC) and the PL with inverse exponential cutoff (PLIEC). Still I follow the principle of parsimony given more than 20 SAR models exist in the literature (Tjørve 2008, 2009, Triantis et al. 2012, Williams et al. 2009) and the excessive computation workload (especially with beta-diversity scaling) had I tested all of the 20+ models, which is hardly necessary with our objectives set for this study. The taper-off parameter ($d$) in both PLEC and PLIEC not only addresses a critique to the traditional power law for overestimating diversity (He & Hubbell 2011), but also preserves the biological interpretations of the scaling parameter (*i.e.*, slope $z$ of SAR) since $d$ is primarily a revision to the other less biologically meaningful parameter $c$ (Tjørve 2009). Furthermore, I propose to define MAD (*maximal accrual diversity*) profile, which can be estimated with the PLEC parameters. I also discuss the possible mechanisms of the DAR scaling such as self-similarity or scale-invariance associated with the power law, and defined a novel *pair-wise diversity overlap* (PDO) metric and the PDO profile, based on the inspirations from existing SAR studies (Harte et al. 1999, 2001, Sizling & Storch 2004, Drakare 2006 and Tjørve et al. 2008).

To the best of our knowledge, this should be the first extension of the SAR to general diversity area scaling beyond species richness level in terms of the Hill numbers. Previously, a few group of researchers, notably Helmus & Ives (2012), Mazel et al. (2014) have successfully extended the SAR to phylogenetic and functional diversities. Their extensions not only verified the applicability of SAR models beyond traditional species richness, but also found important applications in identifying more comprehensive conservation hotspots and predicting the impacts of habitat loss. I expect that our methodological extensions of SAR to general Hill numbers based DAR should not only enrich the theoretical modeling of the diversity scaling in terms of more comprehensive diversity profiles, but also offer important novel insights for their ecological applications. For example, it is well known that species richness alone as a diversity metric, which ignores the equally (if not more) important species abundance and the characteristic of species abundance distribution (SAD), is seriously flawed, and consequently limiting the analysis of diversity scaling to the traditional SAR is unlikely to offer a complete picture of the true diversity changes with area. In addition, the Hill numbers, as metric for diversity profile, have advantages lacked by other existing diversity measures. For instances, the Hill numbers effectively capture the distribution properties of species abundance (rarity *vs*. commonness) by computing the entropy at different orders (non-linearity levels); they are particularly suitable for multiplicative beta-diversity partition (Ellison 2010, Chao et al. 2012, 2014, Jost 2007, which allow us to successfully perform DAR analysis with both alpha- and beta-diversity. These advantages of Hill numbers are naturally carried over to our DAR extensions.

# Materials and Methods

## The American gut microbiome project (AGP) dataset

I use the datasets from the American Gut Project (AGP: http://americangut.org/), which is part of the Earth Micorbiome Project (EMP), and is co-founded by Dr. Rob Knight and Jeff Leach at the University of California, San Diego. The dataset of OTU tables (which are equivalent to the species abundance data of a community in macro-ecology and utilized to test the DAR extensions throughout this article), were rarefied to 10,000 sequence reads per sample computed from the DNA-sequencing data of the 16s-rRNA (v4 region) marker genes from the gut microbiome of 6500 volunteer participants (as of October 2015), was downloaded from the AGP website (https://github.com/biocore/American-Gut/tree/master/data/AG). According to AGP website (http://americangut.org/about/), the protocols used by the AGP project to process the samples and obtain the OTU tables have been extensively tested and benchmarked by Knight Lab at the University of California, San Diego, one of the largest microbiome research labs in the world. I selected the dataset of 1473 healthy Caucasian individuals and excluded the samples from individuals with IBD, diabetes and any other diseases.

## Extending SAR to DAR

I use the following *definitions* and *procedures* to design, perform and interpret DAR modeling analysis with the AGP datasets. To save page space, their detailed descriptions are presented in the online Supplementary Documents (I).

### (*i*) Definitions of alpha and beta diversities

I adopt the Hill numbers to measure both alpha and beta diversities, and *multiplicative partition* of the Hill numbers to define beta diversity.

The Hill numbers, originally introduced as an *evenness* index from economics by Hill (1973), has not received the attention it deserves until recent years. Jost (2007) and Chao et al. (2012) further clarified Hill's numbers for measuring alpha diversity as:

$$^qD = \left(\sum_{i=1}^{S} p_i^q\right)^{1/(1-q)} \qquad (1)$$

where $S$ is the number of species, $p_i$ is the relative abundance of species $i$, $q$ is the order number of diversity.

The Hill number is undefined for $q=1$, but its limit as $q$ approaches to $1$ exists in the following form:

$$^{1}D = \lim_{q \to 1} {^{q}D} = \exp\left(-\sum_{i=1}^{S} p_i \log(p_i)\right) \qquad (2)$$

The parameter $q$ determines the sensitivity of the Hill number to the relative frequencies of species abundances. When $q=0$, the species abundances do not count at all and $^{0}D=S$, *i.e.,* species richness. When $q=1$, $^{1}D$ equal the *exponential* of Shannon entropy, and is interpreted as the number of typical or common species in the community. When $q=2$, $^{2}D$ equal the reciprocal of Simpson index, *i.e.*,

$$^{2}D = \left(1 / \sum_{i=1}^{S} p_i^2\right) \qquad (3)$$

which is interpreted as the number of dominant or very abundant species in the community (Chao et al. 2012). The general interpretation of $^{q}D$ (diversity of order $q$) is that the community has a diversity of order $q$, which is equivalent to the diversity of a community with $^{q}D=x$ equally abundant species.

Recent studies (*e.g.*, Jost 2007, Ellison 2010, Chao et al. 2012, Gotelli & Chao 2013) have advocated the use of multiplicatively defined beta-diversity, rather than additively defined, by partitioning gamma diversity into the product of alpha and beta, in which both alpha ($^{q}D_{\alpha}$) and gamma ($^{q}D_{\gamma}$) diversities are measured with the Hill numbers.

$$^{q}D_{\beta} = {^{q}D_{\gamma}} / {^{q}D_{\alpha}} \qquad (4)$$

This beta diversity ($^{q}D_{\beta}$) derived from the above partition takes the value of $1$ if all communities are identical, the value of $N$ (the number of communities) when all the communities are completely different from each other (there are no shared species). With Jost (2007) words, this beta diversity measures "*the effective number of completely distinct communities.*" In this article, I compute diversities until $q=3$, *i.e.*, to the third order. Note that a series of the Hill numbers at different order $q$ is termed *diversity profile* (Jost 2007, Chao et al. 2012).

### (*ii*) The DAR models and DAR profiles

Since all Hill numbers are in units of species, and in fact, they are referred to as the effective number of species or as species equivalents, intuitively, Hill numbers should follow the same or similar pattern of SAR. I postulate that, similar to the well-known SAR for species *richness* (*i.e.*, the Hill number of order zero, $^{0}D=R$), there exist counterparts for the Hill numbers of general $q$-order, $^{q}D$. I set to investigate the extensions of SAR to general diversity scaling with area (DAR) and further verify our extensions with the AGP dataset.

The basic power function, known as the *power law* (PL) species scaling law widely adopted in SAR study, is extended to describe the general diversity-area relationship (DAR):

$$^qD = cA^z \tag{5}$$

where $^qD$ is diversity measured in the *q-th* order Hill numbers, $A$ is *area*, and $c$ & $z$ are parameters.

I also extend two modified PL models for DAR analysis: power law with exponential cutoff (PLEC) and power law with inverse exponential cutoff (PLIEC), originally introduced to SAR modeling by Plotkin et al. (2000) and Ulrich & Buszko (2003), respectively (also see Tjørve 2009). The PLEC model is:

$$^qD = cA^z \exp(dA), \tag{6}$$

where $d$ is a third parameter and should be negative in DAR scaling models, and $\exp(dA)$ is the exponential decay term that eventually overwhelms the power law behavior at very large value of $A$. The justification for adding the exponential decay term is because both the human body and microbial species inhibited on or in human body are finite, and there should be a taper-off item to reflect the finite size of diversity.

PLIEC is similar to PLEC but with *Sigmoid* shape, rather than convex as PLEC; it is,

$$^qD = cA^z \exp(d/A). \tag{7}$$

Essentially, PLEC and PLIEC can be considered as extensions to parameter $c$, rather than $z$, i.e., $c(x) = c\exp(dx)$ or $c(x) = c\exp(d/x)$, respectively. Therefore, $z$ is assumed to have the similar interpretation as its counterpart in the basic PL. PLEC and PLIEC, however, both behave very differently. The PLEC model asymptotically approaches $cx^z$ as $x$ becomes small, whereas the PLIEC asymptotically approaches $cx^z$ as $x$ becomes large. They were designed to remedy the potentially *unlimited accrual* of species when the area approaches to infinity by introducing a taper-off exponent that may even produce asymptote.

I use the following log-linear transformed equations (8)(9)(10) to estimate the model parameters of Eqn. (5)(6)(7), respectively:

$$\ln(D) = \ln(c) + z\ln(A) \tag{8}$$

$$\ln(D) = \ln(c) + z\ln(A) + dA \tag{9}$$

$$\ln(D) = \ln(c) + z\ln(A) + d/A. \tag{10}$$

I consider the ability to fit all three models (PL, PLEC & PLIEC) in a unified manner—linear transformation—an advantage. I use both linear correlation coefficient ($R$) and *p*-value to judge the goodness of the model fitting. In fact, either of them should be sufficient to judge the suitability of the models to data. An even more important advantage is that the three models preserve the ecological interpretation of the scaling parameter $z$.

Adopting the convention in SAR analysis, the fitted parameter *z* with Eqn. (8) is termed the slope of the power-law DAR, because *z* represents the *slope* of the linearized function in log–log space. However, the *slope* of the DAR as the *tangent* to the curve in the untransformed axes [*i.e.,* the original PL-DAR, Eqn. (5)] is determined by both fitted parameters *z* and *c* as explained in the online Supplementary document. This is a significant advantage of the log-transformed fitting of SAR, and also the primary reason why I adopted the log-log linearized fitting in this study.

I define the relationship between DAR model parameter (*z*) of the traditional PL model and the diversity order (*q*), or *z-q* trend, as the *DAR profile*. It describes the change of diversity scaling parameter (*z*) with the diversity order (*q*), comprehensively. Our definition is obviously inspired by the *diversity profile* of the Hill numbers (Chao et al. 2012, 2014).

### (*iii*) Sampling schemes to fit DAR models

Proper sampling schemes and the accrual of areas are not obvious in our study. I found that Scheiner (2003, 2011) type-III sampling scheme (*i.e., non-contiguous quadrats grid*) is the most appropriate for DAR modeling. Arguments for designing the sampling schemes are provided in the online Supplementary Documents-(I).

Unlike most studies in macro-ecology, where there is often a natural spatial sequence (or arrangement) among the communities sampled, there is not a naturally occurring spatial sequence (arrangement) among the communities of individual subjects from whom AGP samples were obtained. To avoid the potential bias from an arbitrary order of the community samples, I totally permutated the orders of all the community samples under investigation, and then randomly choose 100 (1000 for alpha-DAR) orders of the communities generated from the permutation operation. That is, rather than taking a single arbitrary order for accruing community samples in one-time fitting to the DAR model, I repeatedly perform the DAR model-fitting 100 (1000) times with the 100 (1000) randomly chosen orders. Finally, the averages of the model parameters from the 100 (1000) times of DAR fittings are adopted as the model parameters of the DAR for the set of community samples under investigation.

### (*iv*) The accrual of diversities to fit DAR models

To devise what I believe to be the most appropriate and also natural scheme to accrue diversity, I follow the following three principles. The first is to use the Hill numbers, or what Jost (2007) termed the *true* diversity; the second is to follow the essence of SAR, as captured by the word "accumulation" or "aggregate," *i.e.*, species (diversity) are accumulated for the accrued areas; the third is that the diversity scaling model should be useful for *predicting* diversity at different levels of areas accumulated. I consider these three principles as axioms in traditional SAR and I believe that any extension from SAR to DAR should not violate them. One important advantage for us to stick to the three principles, which are embodied in the traditional SAR theory, is that our new DAR may inherit many of the insights and applications traditional SAR has reveled and offered. The accrual scheme based on the three axioms is described in detail in the online Supplementary Document-(I).

### (*v*) Predicting MAD (Maximal Accrual Diversity) with PLEC-DAR models

The wide application of the traditional SAR in the theory and practice of the global biodiversity conservation sets an excellent precedent for the biomedical applications of the DAR models I build in this study. For example, one may use the DAR models predict the (accumulated) diversities in a human population. In the following, I present one novel application—estimation of the maximal accrual diversity (MAD) of the human microbiome with PLEC model. Among the three DAR models, only PLEC may have a maximum, as derived below based on PLEC model of DAR.

The necessary condition for Eqn. (6) to achieve maximum is its derivative equals zero, *i.e.*,

$$\frac{df(A)}{dA} = \left(^qD\right)' = \left[cA^z \exp(dA)\right]' = 0$$

$$czA^{z-1}\exp(dA) + cA^z \exp(dA)d = 0$$

$$czA^{z-1} + cA^z d = 0 \quad (c \neq 0)$$

$$zA^{z-1} + A^z d = 0$$

$$z + Ad = 0$$

Hence, when

$$A_{max} = -z/d \tag{11}$$

$^qD$ may have a maximum in the following form:

$$Max(^qD) = c\left(-\frac{z}{d}\right)^z \exp(-z) = cA_{max}^z \exp(-z) \tag{12}$$

Eqs. (11) & (12) can be utilized to predict the maximal accrual diversity (MAD) of the human microbiome, whether it is alpha or beta diversity. I define the MAD profile as the relationship between the $D_{max}$ and diversity order $q$, *i.e.*, $D_{max}$–$q$ trend. It is noted that in the above derivation, there are two implicit assumptions: one is that $A_{max}>0$, which requires $z$ and $d$ of different signs, and another is $z>0$, $d<0$. The situation restricted by the first assumption is ecologically meaningless and I can safely eliminate it from consideration because negative accrual ($A_{max}<0$) is not possible. The situation restricted by the second assumption (*i.e.*, $z<0$ & $d>0$) is possible both mathematically and ecologically, but the extreme value is then minimum rather than maximum. In the case of the traditional SAR, the $z<0$ is not justified. However, in general DAR with Hill numbers, $z<0$ is possible at higher diversity orders. In this study, I use the average $z$ and $d$ from 100/1000 times of re-sampling operations, to compute $D_{max}$. In case the average $z$ and $d$ do not satisfy the above two assumptions, I select the valid permutations from 100/1000 re-samplings, compute $D_{max}$ for each valid permutation and then obtain the average $D_{max}$ of the valid permutations.

(*vi*) **The self-similarity property and pair-wise diversity overlap (PDO) profile**

Since diversity measured in Hill numbers are the numbers of *species equivalents*, I expect that the PL-DAR should possess the self-similarity or scale-invariance as SAR has demonstrated (Harte et al. 1999, 2001, Sizling & Storch 2004, Drakare 2006 and Tjørve et al. 2008). Adopting similar derivation process with the SAR, the following properties of PL-DAR can be worked out as follows:

From Eqn. (5), the following equations can be derived:

$$dD/dA = zD/A \tag{13}$$

$$\frac{dD/D}{dA/A} = z. \tag{14}$$

Hence, $z$ is the ratio of *diversity accrual rate* to *area increase rate*.

By setting $A=1$, $S_0 = cA^z = c$, hence $c$ is *the number of species equivalents of diversity in one unit of area,* but not per unit of area because the scaling is nonlinear.

The self-similarity is also known as scale-invariant, which refers to the following mathematical property of the power law:

$$f(\alpha A) = c(\alpha A)^z = \alpha^z f(A) \propto f(A) \tag{15}$$

that is, scaling the argument $A$ (area) by a constant factor $\alpha$, is equivalent to scaling its function proportionally by a constant factor $\alpha^z$. Therefore, all power laws with a particular scaling exponent $z$ are equivalent up to constant factors because each is a scaled version of the others. The scale-invariance is also responsible for the linear relationship after log-transformation of power law [Eqn. (8)], and the resulted straight line on log-log plot is termed the signature of power law. This is another reason I adopted log-log linear transformation fitting of the power law; of course, this is essentially the same argument I argued previously (*i.e.*, the 'slope' argument).

From (15), it is also obvious that:

$$D_{\alpha A}/D_A = \alpha^z \tag{16}$$

where $D_{\alpha A}$ and $D_A$ are the diversity at area size $\alpha A$ and $A$, respectively, $\alpha^z$ is the scaling factor. I omitted diversity order ($q$) to simplify the notation, *e.g.*, $D_A$ in place of $^q D_A$.

Applying log function with the base ($\alpha$) on both sides of (16), there is:

$$\log_\alpha(D_{\alpha A}/D_A) = \log_\alpha \alpha^z = z \log_\alpha \alpha = z \tag{17}$$

It follows that

$$D = cA^{\log_\alpha(D_{\alpha A}/D_A)} \tag{18}$$

If $\alpha=2$, then $z = \log_2(D_{2A}/D_A)$

$$D = cA^{\log_2(D_{2A}/D_A)} \qquad (19)$$

is a special case of (18).

The fraction ($h$) of new diversity due to expansion of $\alpha$ times of original area $A$ can be expressed as:

$$h = (D_{\alpha A} - D_A)/D_A = \alpha^z - 1 \qquad (20)$$

Similarly, the proportion of new diversity in the $j$-th area (of the *same* size) added can be computed with the following equation:

$$h_j = (D_{jA} - D_{(j-1)A})/D_A = j^z - (j-1)^z \qquad (21)$$

Tjørve et al (2008) termed $\alpha$ as *area multiplication rate* and I adopt the same term for DAR, and $h$ is the fraction of *new diversity accumulated as a function* of $z$. When $\alpha=2$, the proportion of new diversity $h = 2^z - 1$, the *diversity overlap* ($g$) of two bordering areas of the same size (computed as the proportion of the new diversity in the second area) is as:

$$g = (2D_A - D_{2A})/D_A = 2 - 2^z \qquad (22)$$

In (22), $g$ is also the scale-invariant overlap because it is the overlap between two areas of the *same* size. If $z=1$, then $g=0$, no overlap; and if $z=0$, $g=1$, totally overlap. In reality, $g$ should between 0 and 1.

Since the *equal size* of *area* assumption is largely true in the case of sampling human microbiome, the parameter z of the PL-DAR can be utilized to estimate the *pair-wise diversity overlap* (PDO), *i.e.*, diversity overlap between two individuals, in the human microbiome with Eqn. (22). Given the range of $g$ is between 0 and 1, I may even use percentage notation to measure pair-wise diversity overlap.

Similar to previous definitions for DAR profile ($z$-$q$ pattern) and MAD profile ($D_{max}$-$q$ pattern), I define PDO profile ($g$-$q$ pattern) as a series of values of the pair-wise diversity overlap metric ($g$) at different diversity order ($q$). The profile comprehensively (at different diversity order or nonlinear level, $q$) captures the average-level, pair-wise overlap (similarity) between two communities in a meta-community setting. Although the $g$ (PDO profile) is simply a precise function of PL-DAR $z$ (DAR profile)[eqn. (22)], the former is far more convenient for measuring community overlap (similarity), which should have more straightforward and intuitive usage.

# Results

Our test of DAR extensions with the AGP dataset consists of two parts: alpha-DAR and beta-DAR modeling, each with three DAR models, PL, PLEC and PLIEC, respectively. I further define DAR, MAD and PDO profiles for the alpha and beta-diversity scaling of the human gut microbiome, respectively. Tables 1-2 list the alpha-DAR models, and Table 3 lists the beta-DAR models. Fig 1 illustrates the DAR-and PDO-profiles for alpha and beta diversities, and Fig 3 illustrates the alpha-MAD profile and beta-MAD profile, respectively.

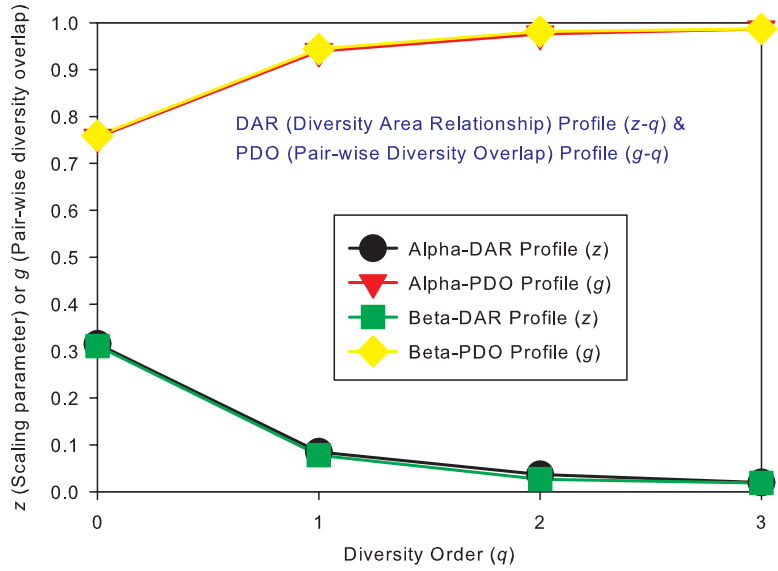

**Fig 1.** The DAR profile and PDO profile for the alpha-diversity and beta-diversity built with the AGP dataset: (*i*) The alpha-DAR profile (*z-q*) and beta-DAR profile (*z-q*) are nearly overlapped, and similarly the alpha-PDO profile (*g-q*) and beta-PDO profile (*g-q*) are nearly overlapped; (*ii*) The DAR profile is monotonically decreasing with diversity order (*q*), and the PDO profile is monotonically increasing with *q*.

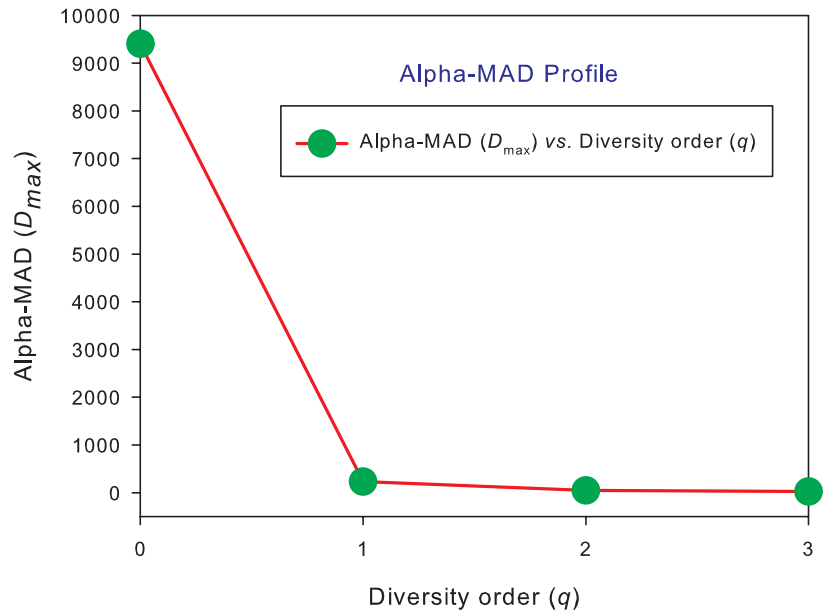

**Fig 2.** The alpha-MAD profile for the alpha-diversity built with the AGP dataset: the MAD profile is monotonically decreasing with diversity order (*q*).

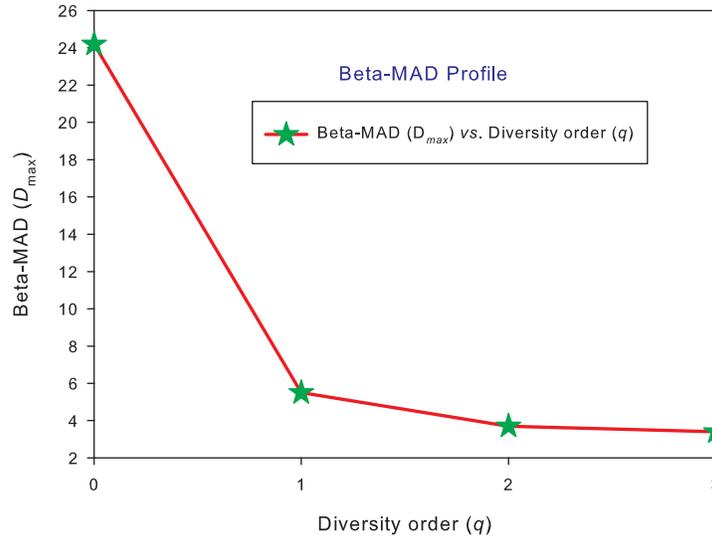

**Fig 3.** The beta-MAD profile for the beta-diversity built with the AGP dataset: the MAD profile is monotonically decreasing with diversity order ($q$).

## Alpha-DAR analysis

Tables 1 and 2 listed the test results of the alpha-DAR modeling with 100 and 1000 times of re-sampling, respectively. Table 3 listed the test results of the beta-DAR modeling with 100 times of sampling. In these tables, I listed: the diversity *order* ($q$) in Hill numbers, model parameters ($z$, ln$c$, $d$), $R$ (linear correlation coefficient), $p$-value measuring the goodness of the model fitting, pair-wise diversity overlap ($g$) and the number of successful fitting of DAR models ($N$). Listed in the last two columns of the PLEC models are the *theoretical maximal accrual diversity* (MAD) ($D_{max}$) and corresponding area accrual ($A_{max}$), predicted with PLEC model [Eqns. (11) & (12)].

**Table 1**. Fitting the alpha-DAR (Diversity Area Relationship) models with 100 times of re-sampling of 1473-Subjects AGP datasets

| Diversity Order & Statistics | | Power Law (PL) | | | | | | PL with Inverse Exponential Cutoff (PLIEC) | | | | | | PL with Exponential Cutoff (PLEC) | | | | | | | |
|---|---|---|---|---|---|---|---|---|---|---|---|---|---|---|---|---|---|---|---|---|---|
| | | $z$ | ln($c$) | $R$ | $p$-value | $g$ | N* | $z$ | $d$ | ln($c$) | $R$ | $p$-value | N* | $z$ | $d$ | ln($c$) | $R$ | $p$-value | N* | $A_{max}$ | $D_{max}$ |
| q=0 | Mean | 0.315 | 6.908 | 0.986 | 0.000 | 0.756 | 100 | 0.291 | -1.493 | 7.067 | 0.996 | 0.000 | 100 | 0.387 | -0.0002 | 6.593 | 0.995 | 0.000 | 100 | 1994 | 9407.4 |
| | Std. Err. | 0.001 | 0.005 | 0.000 | 0.000 | 0.001 | | 0.001 | 0.024 | 0.006 | 0.000 | 0.000 | | 0.001 | 0.0000 | 0.007 | 0.000 | 0.000 | | | |
| | Min | 0.297 | 6.775 | 0.975 | 0.000 | 0.742 | | 0.271 | -2.211 | 6.950 | 0.990 | 0.000 | | 0.347 | -0.0003 | 6.387 | 0.988 | 0.000 | | | |
| | Max | 0.332 | 7.023 | 0.993 | 0.000 | 0.771 | | 0.308 | -0.946 | 7.214 | 0.999 | 0.000 | | 0.429 | -0.0001 | 6.809 | 0.999 | 0.000 | | | |
| q=1 | Mean | 0.085 | 4.849 | 0.789 | 0.000 | 0.939 | 100 | 0.058 | -1.677 | 5.027 | 0.930 | 0.000 | 100 | 0.165 | -0.0002 | 4.504 | 0.900 | 0.000 | 100 | 775 | 229.2 |
| | Std. Err. | 0.002 | 0.014 | 0.009 | 0.000 | 0.001 | | 0.002 | 0.054 | 0.016 | 0.004 | 0.000 | | 0.003 | 0.0000 | 0.018 | 0.006 | 0.000 | | | |
| | Min | 0.044 | 4.475 | 0.529 | 0.000 | 0.900 | | 0.010 | -3.350 | 4.624 | 0.816 | 0.000 | | 0.081 | -0.0004 | 4.100 | 0.654 | 0.000 | | | |
| | Max | 0.138 | 5.130 | 0.943 | 0.000 | 0.969 | | 0.115 | -0.599 | 5.357 | 0.987 | 0.000 | | 0.239 | -0.0001 | 4.935 | 0.988 | 0.000 | | | |
| q=2 | Mean | 0.037 | 3.585 | 0.508 | 0.000 | 0.976 | 90 | 0.014 | -1.207 | 3.740 | 0.763 | 0.000 | 100 | 0.086 | -0.0001 | 3.386 | 0.664 | 0.000 | 99 | 622 | 47.0 |
| | Std. Err. | 0.003 | 0.020 | 0.027 | 0.000 | 0.002 | | 0.003 | 0.061 | 0.023 | 0.014 | 0.000 | | 0.005 | 0.0000 | 0.026 | 0.023 | 0.000 | | | |
| | Min | -0.012 | 3.040 | 0.061 | 0.000 | 0.917 | | -0.047 | -2.764 | 3.125 | 0.431 | 0.000 | | -0.017 | -0.0003 | 2.825 | 0.080 | 0.000 | | | |
| | Max | 0.115 | 3.959 | 0.955 | 0.019 | 1.008 | | 0.102 | 0.035 | 4.191 | 0.971 | 0.000 | | 0.193 | 0.0001 | 3.919 | 0.976 | 0.009 | | | |
| q=3 | Mean | 0.020 | 3.045 | 0.465 | 0.000 | 0.987 | 94 | 0.005 | -0.846 | 3.143 | 0.667 | 0.000 | 100 | 0.052 | -0.0001 | 2.907 | 0.601 | 0.000 | 100 | 586 | 24.3 |
| | Std. Err. | 0.003 | 0.022 | 0.024 | 0.000 | 0.002 | | 0.003 | 0.060 | 0.023 | 0.020 | 0.000 | | 0.005 | 0.0000 | 0.027 | 0.022 | 0.000 | | | |
| | Min | -0.030 | 2.512 | 0.057 | 0.000 | 0.931 | | -0.056 | -2.235 | 2.512 | 0.109 | 0.000 | | -0.064 | -0.0003 | 2.308 | 0.083 | 0.000 | | | |
| | Max | 0.096 | 3.423 | 0.956 | 0.027 | 1.021 | | 0.096 | 0.566 | 3.584 | 0.961 | 0.000 | | 0.169 | 0.0002 | 3.517 | 0.963 | 0.006 | | | |

\* The failed fitting cases (100–N) were removed to compute the statistics of the model parameters, but the major results such as the mean of the model parameters have little differences from the results without removing the failed fitting cases (see Table S1 in Supplementary document (I)). Table S4 in Supplementary document (II) listed the 100 alpha-DAR models from the 100 times of re-sampling.

**Table 2**. Fitting the alpha-DAR (Diversity Area Relationship) models with 1000 times of re-sampling of 1473-Subjects AGP datasets

| Diversity Order and Statistics | | Power Law (PL) | | | | | | PL with Inverse Exponential Cutoff (PLIEC) | | | | | | PL with Exponential Cutoff (PLEC) | | | | | | $A_{max}$ | $D_{max}$ |
|---|---|---|---|---|---|---|---|---|---|---|---|---|---|---|---|---|---|---|---|---|---|
| | | $z$ | $\ln(c)$ | $R$ | $p$-value | $g$ | *N | $z$ | $d$ | $\ln(c)$ | $R$ | $p$-value | *N | $z$ | $d$ | $\ln(c)$ | $R$ | $p$-value | *N | | |
| $q=0$ | Mean | 0.314 | 6.911 | 0.986 | 0.000 | 0.757 | 1000 | 0.290 | -1.508 | 7.071 | 0.996 | 0.000 | 1000 | 0.386 | -0.0002 | 6.598 | 0.995 | 0.000 | 1000 | 2006 | 9413.6 |
| | Std. Err. | 0.000 | 0.002 | 0.000 | 0.000 | 0.000 | | 0.000 | 0.008 | 0.002 | 0.000 | 0.000 | | 0.000 | 0.0000 | 0.002 | 0.000 | 0.000 | | | |
| | Min | 0.291 | 6.771 | 0.954 | 0.000 | 0.739 | | 0.262 | -3.261 | 6.917 | 0.989 | 0.000 | | 0.336 | -0.0003 | 6.295 | 0.978 | 0.000 | | | |
| | Max | 0.335 | 7.081 | 0.995 | 0.000 | 0.777 | | 0.312 | -0.847 | 7.269 | 0.999 | 0.000 | | 0.456 | -0.0001 | 6.866 | 0.999 | 0.000 | | | |
| $q=1$ | Mean | 0.082 | 4.870 | 0.784 | 0.000 | 0.942 | 1000 | 0.054 | -1.714 | 5.051 | 0.935 | 0.000 | 1000 | 0.157 | -0.0002 | 4.542 | 0.890 | 0.000 | 1000 | 780 | 228.6 |
| | Std. Err. | 0.001 | 0.004 | 0.003 | 0.000 | 0.000 | | 0.001 | 0.020 | 0.005 | 0.001 | 0.000 | | 0.001 | 0.0000 | 0.007 | 0.002 | 0.000 | | | |
| | Min | 0.016 | 4.438 | 0.350 | 0.000 | 0.895 | | -0.007 | -5.275 | 4.568 | 0.738 | 0.000 | | 0.022 | -0.0004 | 3.861 | 0.412 | 0.000 | | | |
| | Max | 0.144 | 5.286 | 0.958 | 0.000 | 0.989 | | 0.125 | -0.339 | 5.469 | 0.993 | 0.000 | | 0.290 | 0.0000 | 5.263 | 0.993 | 0.000 | | | |
| $q=2$ | Mean | 0.030 | 3.635 | 0.469 | 0.001 | 0.980 | 948 | 0.009 | -1.224 | 3.775 | 0.749 | 0.000 | 1000 | 0.072 | -0.0001 | 3.454 | 0.638 | 0.000 | 995 | 616 | 46.8 |
| | Std. Err. | 0.001 | 0.007 | 0.008 | 0.000 | 0.001 | | 0.001 | 0.023 | 0.007 | 0.005 | 0.000 | | 0.002 | 0.0000 | 0.010 | 0.007 | 0.000 | | | |
| | Min | -0.068 | 2.942 | 0.052 | 0.000 | 0.905 | | -0.088 | -4.339 | 3.025 | 0.083 | 0.000 | | -0.102 | -0.0004 | 2.570 | 0.079 | 0.000 | | | |
| | Max | 0.131 | 4.301 | 0.956 | 0.047 | 1.046 | | 0.118 | 0.499 | 4.438 | 0.983 | 0.006 | | 0.245 | 0.0002 | 4.403 | 0.966 | 0.010 | | | |
| $q=3$ | Mean | 0.014 | 3.083 | 0.429 | 0.001 | 0.991 | 929 | -0.001 | -0.861 | 3.181 | 0.655 | 0.000 | 997 | 0.039 | -0.0001 | 2.977 | 0.577 | 0.000 | 996 | 561 | 24.2 |
| | Std. Err. | 0.001 | 0.007 | 0.008 | 0.000 | 0.001 | | 0.001 | 0.023 | 0.007 | 0.007 | 0.000 | | 0.002 | 0.0000 | 0.010 | 0.007 | 0.000 | | | |
| | Min | -0.098 | 2.345 | 0.052 | 0.000 | 0.914 | | -0.108 | -3.746 | 2.403 | 0.071 | 0.000 | | -0.153 | -0.0004 | 2.028 | 0.078 | 0.000 | | | |
| | Max | 0.119 | 3.842 | 0.957 | 0.048 | 1.065 | | 0.112 | 0.901 | 3.906 | 0.976 | 0.025 | | 0.213 | 0.0003 | 3.985 | 0.971 | 0.012 | | | |

* The failed fitting cases (1000–$N$) were removed to compute the statistics of the model parameters, but the major results such as the mean of the model parameters have little differences from the results without removing the failed fitting cases (see Table S2 in Supplementary document (I)). Table S5 in Supplementary document (II) listed the 1000 alpha-DAR models from the 1000 times of re-sampling.

From both Tables (1) & (2), I expose the following findings regarding the test of alpha-DAR models:

(i) *The performance of alpha-DAR models*: The number of successful fittings ($N$) shows that at lower diversity order $q=0$ & 1, all three DAR models fitted to the AGP dataset successfully ($p<0.0001$) in 100% of the sampled cases in both 100 and 1000 times of re-sampling operations. At high diversity order $q=2$ & 3, the PLEC and PLIEC succeeded in 99% sampling cases, and both the models performed slightly better than the PL model (90-95%)($p<0.01$). The linear correlation coefficients ($R$) confirmed the finding. For example, with PL model, at lower diversity order, $R$ ranges between 0.94-0.99, and at higher diversity order, $R$ ranges between 0.47-0.51. The decreased goodness-of-fit is expected since the higher-order Hill numbers have relatively stronger nonlinearity. Although either $p$-value or $R$ alone is sufficient to show the model fitting, I present both to show more comprehensive information ($R$ showing the level of linear correlation). I conclude from the above finding that the extension of SAR to alpha-DAR (in the Hill numbers) with three DAR models is fully justified and verified with the AGP dataset, a single largest HMP dataset I am aware of. All three models are sufficient to describe alpha-DAR, and the PL model is preferred if one is in favor of the principle of parsimony. PLIEC performed the best, but PLEC has an advantage over the other two models in predicting the MAD and establishing the MAD profile—$D_{max}$-$q$ pattern. The finding also shows that 100 times of re-sampling operations are enough to deal with the random noise from arbitrarily setting the accrual order of individuals, given the results from both 100 and 1000 times of samplings had little difference.

I now discuss a potential complication arisen from extending SAR to DAR, *i.e.*, negative scaling parameter (z) at higher diversity order *q*=2-3. Table 4 below listed the number of negative z-values or positive *d*-values (to be discussed later) from fitting the three DAR models. The percentages of negative *z* of the three models PL, PLIEC, and PLEC at *q*=2 for alpha-diversity DAR are 11%, 37%, and 5% respectively, and at *q*=3, 30%, 44% and 12%, respectively. Since these percentages numbers were computed from 100 (1000) times of DAR models from re-sampling of the permutation orders of a single dataset, rather than multiple datasets, I consider the negative *z* was largely due to arbitrary ordering for diversity accrual, which is also the very reason why I adopt the average of 100 times of re-sampling. If the average *z* from the 100 (1000) times of re-ordering (re-sampling from total permutations of the 1473 individual in AGP dataset) is positive, I still consider the DAR model for the AGP dataset as positive DAR scaling.

Of course, I need to answer a more fundamental question, are negative z-values justified ecologically? Our answer is yes. This is because at higher diversity orders, unlike species richness, diversity does not necessary rise in an accrued assemblage (community). For example, rare species in individual assemblage may become commoner, rarer or the same level of rareness when the assemblage is pooled together with another assemblage. Consequently, the diversity of the pooled community could be up, down or unchanged. As a side note, as mentioned previously, since parameter *d* in PLEC and PLIEC is an extension to *c*, rather than *z*, parameter *z* should have similar ecological interpretations as in the original PL model. Therefore, I consider negative *z* in the three DAR models as an ecological reality, rather than a mathematical artifact. In the case of AGP dataset, I adopt the average *z* of 100 times re-sampling of the permutation orders because there is not a natural order to accrue the diversity. If there is a natural order for accruing the diversity, that order should be followed to fit the DAR model, and the sign of *z* should be determined by the natural order.

An additional issue, similar to the sign of z, is the sign of *d* in PLEC and PLIEC. In both PLEC and PLIEC, *d* as an exponential cutoff parameter is usually negative. However, when *z*<0, it is possible that *d*>0. This has an implication for computing MAD ($D_{max}$), as in explanation for Eqns. (11) & (12) in previous section on the derivation of MAD. Indeed, as shown in Table 4, in the case of PLEC, negative *z* is always matched with positive *d*.

Yet another interesting finding can be observed from Tables 1, 2, & 4 (also Tables S4-S6), while PLIEC has the best statistical fitting judged from *p*-value and *R*, followed by PLEC and PL, PLEC has the lowest numbers of negative *z*, followed by PL and PLIEC. If I consider negative *z* a potential issue, although which may not be an issue at all as explained previously, PLEC seems to have an advantage of the lowest percentages of negative z-values, besides being able to predict MAD. The advantage of PL model is its simplicity and established ecological interpretations, but it may fail to fit

DAR data at higher diversity orders. Table 4 also suggests that PLIEC has the highest percentage of negative *z*-values, and yet, negative z-values are not matched with positive *d*-values as in the case of PLEC. I am concerned that, although PLIEC has the best statistical fitting, its behavior may be unnecessarily more complicated than the PL and PLEC models. In consideration of the findings discussed above, I recommend the utilizations of PL for DAR profile and PDO profile, and PLEC for MAD profile, at least for the study of human microbiome.

(ii) ***The parameter ranges of alpha-DAR models***: In all three alpha-DAR models, the scaling exponent ($z$) decreases with the increase of the diversity order ($q$). The alpha-DAR profile, *i.e.*, the $z$-$q$ series with the PL model is [0.315, 0.085, 0.037, 0.020]. The counterpart series for PLIEC and PLEC are [0.291, 0.058, 0.014, 0.005], and [0.387, 0.165, 0.086, 0.052] respectively. Hence, the alpha-DAR profile is a monotonically decreasing curve (Fig 1). Since existing literature has not established a systematic range for the diversity-scaling parameter ($z$) beyond species richness, comparison with existing studies is limited to *zero*-order alpha diversity (*i.e.,* SAR). According to Green & Bohannan's (2006) review, the reported SAR exponents in microbes were in the range between 0.019-0.470, but most values were below 0.2 (8 out of 11 studies). Peay et al (2007) reported a range of 0.20-0.23 eukaryotic soil microbes. A major limitation of these early pioneering studies on the testing of SAR with microbes is then low throughput of DNA sequencing technology in detecting bacteria, and consequently the diversity and SAR exponent may be significantly underestimated. Recent studies further confirmed the validity of microbial SAR (*e.g*., Ruff et al. 2015, van der Gast 2013, 2015). As to the range of *z*-value in plants and animals in macro ecology literature, there are many reports but most pointed to a range between *0.2-0.4*. A more recent large-scale investigation with 601 data sets from terrestrial islands by Triantis et al (2012) revealed a full range from 0.064 to 1.312 with 51% fell between 0.2 and 0.4, 25% exceeded 0.4, and an average of $z$=0.321. Our study hence not only falls in the general range, but also happens to be rather close to the average (0.315 vs. 0.321) reported in macro ecology.

(iii) ***Alpha MAD-profile Prediction***: The last two columns in Tables 1 & 2 listed the alpha-MAD profile or $D_{max}$–q predicted by the alpha-DAR PLEC models, that is, $D_{max}$=[9434, 229.7, 47.4, 24.5] (Fig 2) and $A_{max}$=[2028, 802, 969, 1135] for ($q$=0, 1, 2, 3). I consider the prediction of $D_{max}$ series rather reasonable based on the existing reports on species richness in the human gut microbiome (HMP Consortium 2012a, 2012b). Nevertheless, I am somewhat reserved with the estimates of $A_{max}$, the number of individuals ('area') corresponding to the MAD, which seems being influenced by the random noise in the process of area/diversity accrual. This is evidenced by the wide range (*max-min*) of $A_{max}$ in Tables 1 & 2, but the corresponding $D_{max}$ estimates are rather robust as indicated by their rather narrow ranges.

(iv) ***Pair-wise alpha diversity overlap***: Based on the self-similarity property of PL-DAR, I introduce a new metric, pair-wise diversity overlap (PDO) (*g*) and PDO profile, as derived previously. The *g*-series (*q*=0-3) or PDO profile for the alpha-DAR is [0.756, 0.939, 0.976, 0.987] (Fig 1). While the inter-individual (inter-personal) similarity at the species richness level ($q$=0) can be relatively low (0.756 or 75.6%), due to functional redundancy, the similarity at higher diversity levels ($q$>0) should be rather high (94%-99%), which explains the observed monotonically increasing pattern of PDO profile.

(v) **Summary on the alpha-DAR**: I reiterate the following four important findings regarding the alpha-DAR scaling: *First*, extending the SAR to alpha-DAR measured in the Hill numbers is appropriate as verified with the AGP dataset. PL-DAR model is preferred in consideration of its simplicity and established ecological interpretations in the literature. PL-DAR parameter $z$ is the *diversity accrual rate* to *area increase rate* or the slope of the linear-transformed PL model. Parameter $c$ is *the number of species equivalents of diversity in one unit of area* (but not per unit of area) since the scaling is nonlinear. Due to the inter-individual heterogeneity (variability), $c$ may be strongly influenced by the accrual order (what I termed random noise). It is mainly for this reason that I performed 100/1000 times of re-sampling operations and computed the averages from sampling to get the DAR model parameters. I also found that 100 times of sampling is enough to get reliable model parameters. *Second*, the *alpha-DAR profile* for $q$=0-3 is $z$=[0.315, 0.085, 0.037, 0.020], monotonically decreasing with the diversity order ($q$). The parameter ($z$) at species richness level ($q$=0) of AGP not only falls in the range of Triantis et al (2012) meta-analysis, but also approaches to the average they reported in macro-ecology (AMGP=0.315 *vs*. Triantis meta-analysis=0.321). *Third*, the PLEC=DAR model can be harnessed to predict *the alpha-MAD profile* for $q$=0-3, $D_{max}$=[9434, 229.7, 47.4, 24.5]. This is essentially the theoretical maximal accrual diversity of the human gut micorbiome, estimated from the AGP dataset. Fourth, based on the self-similarity property, the *pair-wise diversity overlap* ($g$) between two individual samples (two humans in AGP case) or the *alpha-PDO profile* for $q$=0 to 3 is $g$=[0.756, 0.939, 0.978, 0.987]. This metric is obviously useful for characterizing the *average* pair-wise similarity (dissimilarity) between two human individuals in their gut microbiome diversity. Although other ecological similarity measures (*e.g.*, reviewed in Chao et al. (2014) in the literature may offer similar information, our new metric ($g$) has an advantage that synthesized information from cohorts such as AGP dataset of 1473 individuals.

**Beta-DAR analysis**

Previous alpha-DAR analysis shows that 100 times of sampling operations are large enough to deal with the random noise from area accrual. I then only sampled 100 times to conduct beta-DAR analysis to save computational resources (I observed that the computing load of beta-diversity analysis is nearly 10 times that for alpha-diversity), and the results are listed in Table 3. The symbols in Table 3 are the same as those in previous Tables 1 & 2 of alpha-DAR analysis. From Tables 3, I obtain the following findings regarding the test of beta-DAR models. Overall, the findings from beta-DAR are rather similar to those from alpha-DAR and therefore, I keep the exposition of Table 3 intentionally brief.

**Table 3**. Fitting the beta-DAR (Diversity Area Relationships) models with 100 times of re-sampling of 1473-Subjects AGP datasets

| Diversity Order and Statistics | | Power Law (PL) | | | | | | PL with Inverse Exponential Cutoff (PLIEC) | | | | | | PL with Exponential Cutoff (PLEC) | | | | | | $A_{max}$ | $D_{max}$ |
|---|---|---|---|---|---|---|---|---|---|---|---|---|---|---|---|---|---|---|---|---|---|
| | | $z$ | $\ln(c)$ | $R$ | $p$-value | $g$ | $N^*$ | $z$ | $d$ | $\ln(c)$ | $R$ | $p$-value | $N^*$ | $z$ | $d$ | $\ln(c)$ | $R$ | $p$-value | $N^*$ | | |
| $q=0$ | Mean | 0.311 | 0.971 | 0.990 | 0.000 | 0.759 | 100 | 0.283 | -2.236 | 1.155 | 0.997 | 0.000 | 100 | 0.377 | -0.0002 | 0.683 | 0.996 | 0.000 | 100 | 2056 | 24.2 |
| | Std. Err. | 0.001 | 0.006 | 0.000 | 0.000 | 0.001 | | 0.001 | 0.024 | 0.008 | 0.000 | 0.000 | | 0.001 | 0.0000 | 0.008 | 0.000 | 0.000 | | | |
| | Min | 0.285 | 0.781 | 0.981 | 0.000 | 0.732 | | 0.251 | -2.848 | 0.963 | 0.992 | 0.000 | | 0.338 | -0.0003 | 0.475 | 0.993 | 0.000 | | | |
| | Max | 0.343 | 1.148 | 0.994 | 0.000 | 0.781 | | 0.314 | -1.789 | 1.376 | 1.000 | 0.000 | | 0.417 | -0.0001 | 0.885 | 0.999 | 0.000 | | | |
| $q=1$ | Mean | 0.078 | 1.167 | 0.808 | 0.000 | 0.944 | 100 | 0.048 | -2.429 | 1.367 | 0.951 | 0.000 | 100 | 0.145 | -0.0002 | 0.878 | 0.910 | 0.000 | 100 | 782 | 5.5 |
| | Std. Err. | 0.002 | 0.015 | 0.011 | 0.000 | 0.002 | | 0.003 | 0.048 | 0.017 | 0.003 | 0.000 | | 0.003 | 0.0000 | 0.019 | 0.007 | 0.000 | | | |
| | Min | 0.022 | 0.833 | 0.439 | 0.000 | 0.908 | | -0.011 | -3.742 | 0.963 | 0.871 | 0.000 | | 0.055 | -0.0004 | 0.502 | 0.595 | 0.000 | | | |
| | Max | 0.126 | 1.554 | 0.957 | 0.000 | 0.984 | | 0.107 | -1.412 | 1.806 | 0.987 | 0.000 | | 0.213 | 0.0000 | 1.410 | 0.993 | 0.000 | | | |
| $q=2$ | Mean | 0.027 | 1.112 | 0.577 | 0.000 | 0.981 | 97 | 0.004 | -1.787 | 1.265 | 0.770 | 0.000 | 100 | 0.074 | -0.0001 | 0.908 | 0.703 | 0.000 | 98 | 610 | 3.7 |
| | Std. Err. | 0.004 | 0.029 | 0.024 | 0.000 | 0.003 | | 0.005 | 0.094 | 0.033 | 0.015 | 0.000 | | 0.006 | 0.0000 | 0.035 | 0.021 | 0.000 | | | |
| | Min | -0.073 | 0.545 | 0.072 | 0.000 | 0.919 | | -0.119 | -4.999 | 0.604 | 0.122 | 0.000 | | -0.099 | -0.0004 | 0.177 | 0.105 | 0.000 | | | |
| | Max | 0.112 | 1.826 | 0.937 | 0.010 | 1.050 | | 0.102 | 0.438 | 2.124 | 0.945 | 0.000 | | 0.195 | 0.0002 | 1.823 | 0.967 | 0.000 | | | |
| $q=3$ | Mean | 0.019 | 1.081 | 0.555 | 0.000 | 0.987 | 99 | -0.001 | -1.534 | 1.209 | 0.701 | 0.000 | 100 | 0.061 | -0.0001 | 0.883 | 0.688 | 0.000 | 97 | 605 | 3.4 |
| | Std. Err. | 0.005 | 0.037 | 0.024 | 0.000 | 0.004 | | 0.006 | 0.125 | 0.043 | 0.017 | 0.000 | | 0.009 | 0.0000 | 0.047 | 0.021 | 0.000 | | | |
| | Min | -0.123 | 0.404 | 0.060 | 0.000 | 0.914 | | -0.178 | -6.124 | 0.478 | 0.205 | 0.000 | | -0.192 | -0.0005 | 0.040 | 0.129 | 0.000 | | | |
| | Max | 0.119 | 1.984 | 0.923 | 0.020 | 1.082 | | 0.107 | 1.008 | 2.389 | 0.933 | 0.000 | | 0.209 | 0.0004 | 2.272 | 0.958 | 0.000 | | | |

\* The failed fitting cases (100–$N$) were removed to compute the statistics of the model parameters, but the major results such as the mean of the model parameters have little differences from the results without removing the failed fitting cases (see Table S3 in Supplementary document (I)). Table S6 in Supplementary document (II) listed the 100 beta-DAR models from the 100 times of re-sampling.

**(i)** *The performance of beta-DAR models*: The goodness-of-fittings of the three DAR models (PL, PLEC, and PLIEC) to the beta-diversity scaling with the AGP dataset are even slightly better than to the alpha-diversity scaling. For example, the minimum percentage of successfully beta-DAR models is 93% in Table 3, compared with 90% in Tables 1 & 2. The minimum of average *R* (linear correlation coefficients) in beta-DAR models (Table 3) is 0.555, higher than that of 0.465 in alpha-DAR models (Table 1). Therefore, beta-diversity scaling can be modeled with the same mathematical functions as alpha-diversity scaling models. To the best of our knowledge, this is the first systematic modeling of the scaling of beta diversity in the Hill numbers.

Similar to the previous alpha-DAR model, I also counted the negative z-values when the three DAR models were fitted to beta-diversity scaling and the results are listed in Table 4 (the same table as for alpha-DAR). The percentages of negative *z* of the three models PL, PLIEC, and PLEC at *q*=2 for beta-diversity DAR are 24%, 40%, and 14% respectively, and at *q*=3, 29%, 42% and 22%, respectively. These percentages are somewhat higher than their alpha-diversity counterparts discussed previously, but our explanations and conclusions are the same as those previously summarized and recommended for the alpha-diversity scaling.

**(ii)** *The parameter ranges of beta-DAR models*: The beta-DAR profile, *i.e.*, the *z*-*q* series with the PL for beta diversity scaling is beta-*z*=[0.311, 0.078, 0.027, 0.019] (Fig 1). This series is rather close to that for alpha-DAR model, which is alpha-*z*=[0.315, 0.085, 0.037, 0.020]. Overall, the scaling patterns for both alpha-DAR and beta-DAR are rather similar. Since existing literature has not established a systematic range for the beta-diversity scaling, there are no existing studies with which I can compare the range of scaling parameters.

(iii) ***Beta MAD-profile Prediction***: The beta MAD profile predicted by the beta-DAR PLEC models, that is, $beta\text{-}D_{max}$=[24.3, 5.5, 3.7, 3.4] (Fig 3) and $beta\text{-}A_{max}$=[2123, 848, 818, 953] for ($q$=0, 1, 2, 3). This beta $D_{max}$-$q$ series is orders of magnitude smaller than its alpha counterpart, which is $alpha\text{-}D_{max}$=[9434, 229.7, 47.4, 24.5], although both the $q\text{-}A_{max}$ series are rather close to each other. The magnitudes of differences in $D_{max}$ between alpha and beta diversity scaling are, of course, expected because the values of alpha and beta diversities are simply at rather different magnitudes.

(iv) ***Pair-wise beta diversity overlap***: Similar to pair-wise alpha diversity overlap, I obtained the $g$-series ($q$=0-3) or PDO profile for beta-DAR is $beta\text{-}g$=[0.759, 0.944, 0.981, 0.987] (Fig 1), which is rather close to that for the alpha-DAR, *i.e.*, $alpha\text{-}g$=[0.756, 0.939, 0.976, 0.987]. This indicates that while the values of alpha diversity and beta-diversity are at different orders of magnitudes, the degree (level) of their pair-wise diversity overlaps is essentially independent of the type of diversity measure adopted (alpha or beta).

(v) ***Summary on the beta-DAR***: When measured in the Hill numbers, the beta-diversity follows the same scaling law as the alpha-diversity does. Indeed, both alpha-DAR and beta-DAR follow the same scaling law as the traditional SAR does. This finding should be expected if I realize that all Hill numbers (either for measuring alpha, beta, or gamma diversities) are in units of *species* (or as *species equivalents*), and measure the *effective number of species*. Indeed, it was this fundamental property of the Hill numbers that motivated us to extend SAR to general DAR. In other words, SAR is a special case of DAR when the diversity order is set to zero (*i.e.*, species richness when $q$=0). The tests with the AMGP dataset verified our postulation that motivated this study.

Table 4. The percentages of negative $z$-values or positive $d$-values in the DAR models with 100(1000) times of re-sampling from the random permutations of 1473 individuals in the AGP datasets

| Diversity Order | Model | Alpah-DAR (100 times) | | Alpha-DAR (1000 times) | | Beta-DAR (100 times) | |
|---|---|---|---|---|---|---|---|
| | | %Negative $z$ | %Positive $d$ | %Negative $z$ | %Positive $d$ | %Negative $z$ | %Positive $d$ |
| $q$=0 | PL | 0 | NA | 0 | NA | 0 | NA |
| | PLIEC | 0 | 0 | 0 | 0 | 0 | 0 |
| | PLEC | 0 | 0 | 0 | 0 | 0 | 0 |
| $q$=1 | PL | 0 | NA | 0 | NA | 0 | NA |
| | PLIEC | 0 | 0 | 0.7 | 0 | 4.0 | 0 |
| | PLEC | 0 | 0 | 0 | 0 | 0 | 0 |
| $q$=2 | PL | 11.1 | NA | 13.3 | NA | 23.7 | NA |
| | PLIEC | 37.0 | 2.0 | 39.8 | 2.50 | 40.0 | 1.0 |
| | PLEC | 5.00 | 5.0 | 11.3 | 11.3 | 13.7 | 13.7 |
| $q$=3 | PL | 29.8 | NA | 35.1 | NA | 29.3 | NA |
| | PLIEC | 44.0 | 8.0 | 53.8 | 11.2 | 42.0 | 9.0 |
| | PLEC | 12.0 | 12.0 | 22.5 | 22.5 | 21.6 | 21.6 |

# Discussion

Multiple mechanisms have been proposed to explain SAR, including more individuals (also known as passive sampling, random placement, rarefaction effect, sampling effect, *etc*), environmental heterogeneity (spatial or temporal), dispersal limitations, population dynamics, niche-based interactions, biotic interactions, multiple species pools, meta-population theory, and self-similarity (*see* reviews by White et al. 2006, Scheiner et al. 2011). In spite of the extensive studies in macro-ecology, little direct experimental evidence exists in the literature to prove or reject those proposed mechanisms. Unlike many physical laws whose mechanisms can be theoretically derived and experimentally verified, ecological laws are usually established inductively by the accumulation of experimental data. Although the accumulated ecological data may establish the validity of an ecological law, the data that can *directly* determine or reveal the mechanism are frequently difficult to collect. Due to this limitation, meta-analysis is often used to investigate the factors (variables) that may affect ecological law. In the case of SAR, quite a few excellent meta-analysis or similar synthesis (not necessarily followed meta-analysis procedure strictly) studies exist (*e.g.*, Drakare et al. 2006). But the results of meta-analysis usually only identify the factors that significantly affect ecological laws (SAR), still may not offer direct evidence to support or reject a specific mechanism hypothesis underlying the law because the complex interaction among the factors are usually hard to consider in meta-analysis, which may play an important role in controlling the behavior of ecosystem (or community). This somewhat unique property of ecological laws also explains why I cannot offer definite conclusion on the mechanism underlying the DAR of the human microbiome. For example, Drakare et al. (2006) meta-analysis with 794 SAR studies reported in major ecological journals revealed that SAR are significantly influenced by variables determining sampling schemes, the spatial scale, and the types of organisms or habitats involved. Those meta-analyses on SAR also offered important insights on the model selection (more than 20 SAR models have been proposed, tested, and evaluated) and other important issues (Tjørve 2009, Triantis et al. 2012, Williams et al. 2009). Our study benefits enormously, especially on the study design including the model selection and fitting, choice of sampling scale (unit), accrual scheme, from the insights and recommendations reported in those existing meta-analyses. Even with these efforts, like many other SAR studies, I could not escape from the general limitation involved in the research of ecological laws.

As demonstrated in previous sections, I systematically extend the traditional SAR relationships to their counterparts of DAR relations for both alpha and beta-diversity, based on the known most appropriate diversity metrics—the Hill numbers. These extensions not only enrich our tools for investigating the biogeography of ecological communities and ecosystems in general, but also significantly expand and deepen our understanding of the biogeographic properties of the human microbiome (*e.g.*, Hanson et al. 2012, Oh et al. 2014), such as its spatial heterogeneity, which in turn is of significant importance for personalized medicine (Ma *et al*. 2011). This is because spatial heterogeneities, translated into individualities, may influence the diagnosis and treatment of microbiome-associated diseases (Ma et al. 2011). The DAR models are likely to offer important guidelines for conserving arguably the most important biodiversity to our health—the diversity of our gut microbiome (O'Doherty et al. 2014). Finally, although I demonstrated the DAR extensions with microbes, the conclusions should also hold in the native *area* of SAR or in macro-ecology.